# Pseudorandomness and Combinatorial Constructions

Luca Trevisan[*]

September 20, 2018

**Abstract**

In combinatorics, the probabilistic method is a very powerful tool to prove the existence of combinatorial objects with interesting and useful properties. Explicit constructions of objects with such properties are often very difficult, or unknown. In computer science, probabilistic algorithms are sometimes simpler and more efficient than the best known deterministic algorithms for the same problem.

Despite this evidence for the power of random choices, the computational theory of pseudorandomness shows that, under certain complexity-theoretic assumptions, every probabilistic algorithm has an efficient deterministic simulation and a large class of applications of the the probabilistic method can be converted into explicit constructions.

In this survey paper we describe connections between the conditional "derandomization" results of the computational theory of pseudorandomness and unconditional explicit constructions of certain combinatorial objects such as error-correcting codes and "randomness extractors."

## 1 Introduction

### 1.1 The Probabilistic Method in Combinatorics

In extremal combinatorics, the *probabilistic method* is the following approach to proving existence of objects with certain properties: prove that a random object has the property with positive probability. This simple idea has beem amazingly successful, and it gives the best known bounds for most problems in extremal combinatorics. The idea was introduced (and, later, greatly developed) by Paul Erdős [18], who originally applied it to the following question: define $R(k,k)$ to be the minimum value $n$ such that every graph on $n$ vertices has either an independent set of size at least $k$ or a clique of size at least $k$;[1] It was known that $R(k,k)$ is finite and that it is at most $4^k$, and the question was to prove a lower bound. Erdős proved that a random graph with $2^{k/2}$ vertices has a positive probability of having no clique and no independent set larger than $k$, and so $R(k,k) \geq 2^{k/2}$. The method, of course, gives no indication of how to actually construct a large graph with no small clique and no small independent set. Remarkably, in the past 60 years, there has been no asymptotic improvement to Erdős's lower bound and, perhaps more significantly, the best *explicit construction* of a graph without a clique of size $k$ and without an independent set of size $k$ has only about $k^{\log k}$ vertices [20], a bound that has not improved in 25 years.

---

[*]Computer Science Division, U.C. Berkeley. `luca@cs.berkeley.edu`. The author is supported in part by NSF grant CCF 0515231.

[1]Here by "graph" we mean an undirected graph, that is, a pair $G = (V, E)$ where $V$ is a finite set of *vertices* and $E$ is a set of pairs of elements of $E$, called *edges*. A clique in a graph $G = (V, E)$ is a set $C \subseteq V$ of vertices such that $\{u, v\} \in E$ for every two vertices $u, v \in C$. An independent set is a set $I \subseteq V$ of vertices such that $\{u, v\} \notin E$ for every two vertices $u, v \in I$.

Shannon [59] independently applied the same idea to prove the existence of encoding schemes that can optimally correct from errors in a noisy channel and optimally compress data. The entire field of information theory arose from the challenge of turning Shannon's non-constructive results into algorithmic encoding and decoding schemes. We will return to the problem of encodings for noisy channels in Section 2. Around the same time, Shannon [60] applied the probabilistic method to prove the existence of boolean function of exponential circuit complexity (see Section 4). Proving that certain specific boolean functions (for example, satisfiability of boolean formulae or 3-colorability of graphs) require exponential size circuits is a fundamental open problem in computational complexity theory, and little progress has been made so far.

The probabilistic method has found countless applications in the past 60 years, and many of them are surveyed in the famous book by Alon and Spencer [4].

For most problems in extremal combinatorics, as well as in information theory and complexity theory, probabilistic methods give the best known bound, and explicit constructions either give much worse bounds, or they give comparable bounds but at the cost of technical tours de force.

## 1.2 Probabilistic Methods in Computer Science

In computer science, an important discovery of the late 1970s was the power of *probabilistic algorithms*. The most famous (and useful) of such algorithms are probably the polynomial time probabilistic algorithms for testing primality of integers [65, 47]. In these algorithms, one looks for a "certificate" that a given number $n$ is composite; such a certificate could be for example an integer $a$ such that $a^n \not\equiv a \pmod{n}$, or four distinct square roots $\pmod{n}$ of the same integer. Rabin, Solovay and Strassen [65, 47] proved that there is a good chance of finding such certificates just by picking them at random, even though no efficient method to deterministically construct them was known.[2] Other two important and influential algorithms were discovered around the same time: an algorithm to test if two implicitly represented multivariate polynomials are identical [85, 56] (evaluate them at a random point chosen from a domain larger than the degree) and an algorithm to check if two vertices in a graph are connected by a path [2] (start a random walk at the first vertex, and see if the second vertex is reached after a bounded number of steps).[3]

A different type of probabilistic algorithms was developed starting in the late 1980s with the work of Sinclair and Jerrum [61]. These algorithms solve approximate "counting" problems, where one wants to know the number of solutions that satisfy a given set of combinatorial constraints. For example, given a bipartite graph, one would like to know, at least approximately, how many perfect matching there are.[4] Sinclair and Jerrum introduced an approach based on a reduction to the problem of approximately sampling from the uniform distribution of all possible solutions. Since the latter problem involves randomness in its very definition, this approach inevitably leads to probabilistic algorithms.

---

[2]Note the similarity with the probabilistic method.

[3]The algorithm of Aleliunas et al. [2] broke new grounds in term of *memory* use, not running time. It was already known that the Depth-First-Search algorithm could be used to solve the problem using linear time and a *linear* amount of memory. The random walk algorithm, however, needs only $O(\log |V|)$ bits of memory, and exponential improvement.

## 1.3 The Computational Theory of Pseudorandomness

In light of such algorithmic results, it was initially conjectured that probabilistic algorithms are strictly more powerful than deterministic ones and that, for example, there exist problems that can be solved probabilistically in polynomial time but that cannot be solved in polynomial time using deterministic algorithms.

This belief has been overturned by developments in the computational theory of pseudorandomness. The theory was initiated by Blum, Goldwasser, Micali and Yao [22, 9, 84], with the motivation of providing sound foundations for cryptography. From the very beginning, Yao [84] realized that the theory also provides conditional *derandomization* results, that is, theorems of the form

> "if assumption X is true, then every problem that can be solved by a probabilistic polynomial time algorithm can also be solved by a deterministic algorithm of running time Y."

Yao [84] showed that we can take X to be "there is no polynomial time algorithm that on input a random integer finds its prime factorization"[5] and Y to be "time $2^{n^\epsilon}$ for every $\epsilon > 0$."

An important project in complexity theory in the 1980s and 1990s was to strengthen Y to be "polynomial time" with a plausible X. The goal was achieved in a landmark 1997 paper by Impagliazzo and Wigderson [31], building on a considerable body of previous work.

At a very high level, the Impagliazzo-Wigderson result is proved in two steps. The first step (which is the new contribution of [31]) is a proof that an assumption about the worst-case complexity of certain problems implies a seemingly stronger assumption about the average-case complexity of those problems. A result of this kind is called an *amplification of hardness* result, because it "amplifies" a worst-case hardness assumption to an average-case one. The second step, already established ten years earlier by Nisan and Wigderson [44] is a proof that the average-case assumption suffices to construct a certain very strong pseudorandom generator, and that the pseudorandom generator suffices to simulate deterministically in polynomial time every polynomial time probabilistic algorithm.

In conclusion, assuming the truth of a plausible complexity-theoretic assumption, every polynomial time probabilistic algorithm can be "derandomized," including the approximation algorithms based on the method of Sinclair and Jerrum. Furthermore, under the same assumption, a large class of applications of the probabilistic method in combinatorics can be turned into explicit constructions.

---

[4]A bipartite graph is a triple $G = (U, V, E)$ where $U, V$ are disjoint sets of vertices and $E \subseteq U \times V$ is a set of edges. A perfect matching is a subset $M \subseteq E$ such that for every $u \in U$ there is precisely one $v \in V$ such that $(u, v) \in M$, and vice versa.

[5]More generally, Yao showed that $X$ can be "one-way permutations exist," see 5 for more details. The assumption about integer factorization implies the existence of one-way permutations, provided that we restrict ourselves to "Blum integers."

We give some more details about the Impagliazzo Wigderson theorem in Section 6. The reader is also referred to the excellent survey paper of Impagliazzo [28] in the proceedings of the last ICM.

It is remarkable that many of the "predictions" coming from this theory have been recently validated unconditionally: Agrawal et al. [1] have developed a deterministic polynomial time algorithm for testing primality and Reingold [53] has developed a deterministic $O(\log n)$ memory algorithm for undirected graph connectivity. One can read about such developments elsewhere in these proceedings.

Here is an example of a question that is still open and that has a positive answer under the complexity-theoretic assumption used in the Impagliazzo-Wigderson work:

- Is there a deterministic algorithm that, on input an integer $n$, runs in time polynomial in $\log n$ and return a prime between $n$ and $2n$?

## 1.4 When Randomness is Necessary

Suppose that, in a distant future, someone proves the assumption used in the Impagliazzo-Wigderson work, so that we finally have an unconditional polynomial time derandomization of all probabilistic algorithm. Would this be the end of the use of randomness in computer science? The answer is no, for at least two reasons.

One reason is that such derandomization would probably not be practical. At a broad qualitative level, we consider polynomial-time algorithms as "efficient" and super-polynomial-time algorithms as "inefficient," and then such a result would establish the deep fact that "efficient" probabilistic algorithms and "efficient" deterministic algorithms have the same power. If the derandomization, however, causes a considerable (albeit polynomial) slow-down, and if it turns a practical probabilistic algorithm into an impractical deterministic one, then the probabilistic algorithm will remain the best choice in applications.

A more fundamental reason is that there are several applications in computer science where the use of randomness is *unavoidable*. For example, consider the task of designing a secure cryptographic protocol in a setting where all parties behave deterministically.

These observations lead us to consider the problem of generating randomness to be used in probabilistic algorithms, cryptographic protocols, and so on. Such generation begins by measuring a physical phenomenon that is assumed to be unpredictable (such as a sequence of physical coin flips) and that will be called a *random source* in the following. Typically, one has access to random sources of very poor quality, and converting such measurements into a sequence of independent and unbiased random bits is a difficult problem. In Section 8 we discuss various approaches and impossibility results about this problem, leading to the definition of *seeded randomness extractor* due to Nisan and Zuckerman [86, 45].

Seeded randomness extractors have an amazing number of applications in computer science, often completely unrelated to the original motivation of extracting random bits from physical sources. They are related to hash functions, to pseudorandom graphs, to error-correcting codes and they are useful in complexity theory to prove, among other things, negative results for the approximability of optimization problems.

Ironically, the problem of generating high-quality random bits for cryptographic application is not satisfactorily solved by seeded randomness extractor (even though it was the original motivation for the research program that led to their definition). *Seedless* randomness extractors are needed for such application, and their theory is still being developed.

## 1.5 Connections

So far, we have discussed (i) the power of probabilistic methods, (ii) the *conditional* results proving that all probabilistic algorithms have a polynomial-time derandomization under complexity assumption, and (iii) the use of seeded randomness extractor to *unconditionally* run probabilistic algorithms in a setting in which only a weak source of randomness is available.

In Section 8 we describe a recently discovered *connection* between (ii) and (iii) and, more generally, between *conditional* results proved in the computational theory of pseudorandomness and *unconditional* explicit constructions of combinatorial objects.

One connection is between error-correcting codes and "hardness amplification" result. This connection has led to the application of coding-theoretic techniques in the study of average-case complexity. It is also possible to use complexity-theoretic techniques to build error-correcting codes, but so far this approach has not been competitive with previously known coding-theoretic techniques.

The second connection is between pseudorandom generators and seeded randomness extractors. This connection has led to improvements in both setting.

Various impossibility results are known for error-correcting codes and randomness extractors. Via these connections, they imply impossibility results for hardness amplification and conditional derandomization. In Section 7 we discuss approaches to sidestep these negative results.

## 2 Pseudorandom Objects: Codes and Graphs

In this section we introduce two examples of very useful combinatorial objects whose existence easily follows from the probabilistic method: error-correcting codes and expander graphs. Explicit constructions of such objects are also known.

## 2.1 Error-Correcting Codes

Consider the process of picking a random set $S \subseteq \{0,1\}^n$ of size $2^k$, $k < n$. If, say, $k = n/2$, then it is easy to show that there is an absolute constant $\delta > 0$ such that, with high probability, every two elements $u, v \in S$ differ in at least $\delta n$ coordinates. By a more careful estimate, we can also see that there is an absolute constant $c$ such that, for every $\epsilon > 0$, it is likely that every two elements of $S$ differ in at least $(1/2 - \epsilon)n$ coordinates with high probability, provided $k \leq c\epsilon^2 n$. For reasons that will be clear shortly, let us change our perspective slightly, and consider the (equivalent) process of picking a random injective function $C : \{0,1\}^k \to \{0,1\}^n$: clearly the same bounds apply.

For two strings $u, v \in \{0,1\}^n$, the Hamming distance between $u$ and $v$ (denoted $d_H(u,v)$) is the number of coordinates where $u$ and $v$ differ, that is

$$d_H(u,v) := |\{i : u_i \neq v_i\}| \tag{1}$$

**Definition 2.1 (Error-correcting code).** *We say that $C : \{0,1\}^k \to \{0,1\}^n$ is a $(n,k,d)$-code if for every two distinct $x, y \in C$, $d_H(C(x), C(y)) \geq d$.*

This concept is due to Hamming [25]. Error-correcting codes are motivated by the following scenario. Suppose we, the *sender*, have a $k$-bit message $M \in \{0,1\}^k$ that we want to transmit to a *receiver* using an unreliable *channel* that introduces errors, and suppose we have a $(n,k,d)$-code $C$. Then we can compute $c = C(M)$ and transmit $c$ over the channel. The receiver gets a string $c'$, which is a corrupted version of $c$, and looks for the message $M'$ that minimizes $d_H(C(M'), c')$. If the channel introduces fewer than $d/2$ errors, then the receiver correctly reconstructs $M$.[6]

Keeping this application in mind, for every given $k$, we would like to construct $(n,k,d)$ codes where $d$ is as large as possible (because then the receiver can tolerate more errors) and $n$ is as small as possible (so that we do not have to communicate a very long encoding). Furthermore, we would like $C()$ and the decoding procedure run by the receiver to be computable by efficient algorithms.

One trade-off between the parameters is that $\frac{d}{n} \leq \frac{1}{2} + o_k(1)$. Keeping in mind that the number of errors that can be corrected is at most $d/2$, this means that the receiver can correctly reconstruct the message only if the number of errors is at most $\left(\frac{1}{4} + o_k(1)\right)n$.

It is possible to do better if we are willing to settle for the notion of "list-decodability," introduced by Elias [17].

**Definition 2.2 (List-decodable code).** *We say that $C : \{0,1\}^k \to \{0,1\}^n$ is $(L, \delta)$-list decodable if for every $u \in \{0,1\}^n$,*
$$|\{x \in \{0,1\}^k : d_H(C(x), u) \leq \delta n\}| \leq L$$

Here the idea is that we send, as before, the encoding $C(M)$ of a message $M$. The receiver gets a string $u$ and computes the *list* of all possible messages $M'$ such that $d_H(C(x), u) \leq \delta n$. If $C$ is a $(L, \delta)$-code, then the list is guaranteed to be of length at most $L$, and if the channel introduces at most $\delta n$ errors then our message is guaranteed to be in the list.

Using the probabilistic method, it is easy to show the existence of $(L, 1/2 - \epsilon)$-list decodable codes $C : \{0,1\}^k \to \{0,1\}^n$ for every $k$ and $\epsilon$, where $n = O(k\epsilon^{-2})$ and $L = O(\epsilon^{-2})$. It was also known how to define *efficiently encodable* codes with good (but not optimal) parameters. It took, however, 40 years until Sudan [66] defined the first *efficient list-decoding* algorithm for such codes. Sudan's algorithm suffices to define $(\epsilon^{-O(1)}, 1/2 - \epsilon)$-list decodable codes $C : \{0,1\}^k \to \{0,1\}^n$ with $n = (k/\epsilon)^{O(1)}$ for every $k, \epsilon$, and the codes are encodable and list-decodable in time polynomial in $n$. This means that even if the channel introduces close to $n/2$ errors, it is still possible for the receiver to gain considerable information about the message. (Namely, the fact that the message is one out of a small list of possibilities.) Other list-decoding algorithms are now known, but they are beyond the scope of this survey. See Sudan's survey [67], Guruswami's thesis [23] and two recent breakthrough papers [46, 24].

---

[6]It is also easy to see that this analysis is tight. If there are two messages $M, M'$ such that $d_H(C(M), C(M')) = d$, and we send $M$, then it is possible that even a channel that introduces only $d/2$ errors can fool the receiver into thinking that we sent $M'$.

## 2.2 Expander Graphs

Consider the process of picking at random a graph according to the $G_{n,\frac{1}{2}}$ distribution.(The $G_{n,\frac{1}{2}}$ distribution is the uniform distribution over the set of $2^{\binom{n}{2}}$ graphs over $n$ vertices.) A simple calculation shows that for every two disjoint sets of vertices $A, B$ there are $\left(\frac{1}{2} \pm o_n(1)\right)|A||B|$ edges with one endpoint in $A$ and one endpoint in $B$. Chung, Graham and Wilson [15] call a family of graphs satisfying the above properties a family of *quasi-random* graphs, and prove that six alternative definitions of quasi-randomness are all equivalent. Explicit constructions of quasi-random graphs are known, and the notion has several applications in combinatorics. (See the recent survey paper by Krivelevich and Sudakov [37].) Consider now a process where we randomly generate an $n$-vertex graph where every vertex has degree at most $d$ (think of $d$ as a fixed constant and $n$ as a parameter), for example, consider the process of picking $d$ perfect matchings and then taking their union. Then it is possible to show that for every two disjoint sets of vertices $A, B$ there are $(1 \pm o_{n,d}(1)) d\frac{|A||B|}{n}$ edges with one endpoint in $A$ and one endpoint in $B$. (Families of) graphs with this property are called *expanders*, and they have several applications in computer science. To gain a sense of their usefulness, imagine that an expander models a communication network and note that if $o(dn)$ edges are deleted, the graph still has a connected component with $(1-o(1))n$ vertices. Furthermore, expander graphs have small diameter, it is possible to find several short edge-disjoint paths between any two vertices, and so on. There are other possible definitions of expanders, which are related but not equivalent. In one possible (and very useful) definition, expansion is measured in terms of the *eigenvalue gap* of the *adjacency matrix* of the graph (see e.g. the discussion in [37]). For this definition, Lubotzky, Phillips and Sarnak [41] provide an optimal explicit construction. Another possible measure is the *edge expansion* of the graph. Optimal explicit constructions for this measure are not known, but considerable progress is made in [13, 3].

# 3 Randomness Extractor

Randomness extractors are procedures originally designed to solve the problem of generating truly random bits. As we will see, randomness extractors can be seen as a sort of pseudorandom graphs, they can be constructed using techniques from the field of pseudorandomness, and they are tightly related to constructions error-correcting codes, expanders and other random-like combinatorial objects.

## 3.1 Generating Random Bits

In order to generate random bits in practice, one starts by measuring a physical phenomenon that is assumed to contain randomness.[7] For example, in many computer systems one starts by collecting statistics on the user's keystrokes or mouse movement, or on the latency time of disk access, and so on. This raw data, which is assumed to contain some amount of entropy, is then passed to a "hash function," and the output of the function is assumed to be a sequence of truly random bits. Such systems, widely used in practice, are typically not validated by any rigorous analysis.

In a mathematical modeling of this situation, we have a random variable $X$, representing our physical measurement, ranging, say, over $\{0,1\}^n$. We would like to construct a function $Ext: \{0,1\}^n \to \{0,1\}^m$ such that, by making as little assumptions on $X$ as possible, we can prove that $Ext(X)$ is distributed uniformly over $\{0,1\}^m$, or at least it is approximately so.

---
[7]We will not get into the physical and philosophical problems raised by such assumption.

Von Neumann [82] studied a version of this problem where $X$ is a sequence of independent and identically distributed biased coin tosses. The independence assumption is crucially used. The general problem was extensively studied in computer science in the 1980s [55, 80, 79, 78, 14, 16]. Notably, the goal was to define a single function $Ext$ that would work for as large as possible a class of distributions $X$. An early conclusion was that the extraction problem is impossible [55], as defined above, even if just very weak forms of dependencies between different bits are allowed in the distribution of $X$. Two approaches have been considered to circumvent this impossibility.

1. One approach is to consider a model with a *small* number of mutually independent random variables $X_1, \ldots, X_k$, each satisfying weak randomness requirements. This line of work, initiated in [55, 80, 14], saw no progress for a long time, until recent work by Barak et al. [7] made possible by a breakthrough in additive combinatorics [12, 36]. This is now a very active area of research [8, 53, 11, 87, 48] with connections to other areas of combinatorics.

2. The other approach, initiated in [79], is to stick to the model of a single sample $X$ and to consider the following question: suppose we have a randomized algorithm $A$ (that is correct with high probability given the ability to make truly random choices) and suppose we have an input $x$. Can we efficiently find what is the most probable output of $A(x)$?

## 3.2 The Definition of Randomness Extractors

To formalize approach (2), it is convenient to think of a probabilistic algorithm $A()$ as having two inputs: a "random" input $r$ and a "regular" input $I$. We say that "$A$ computes a function $f$ with high probability" if, for every $I$

$$\mathbb{P}[A(r,I) = f(I)] \geq .9$$

where the probability is taken with respect to the uniform distribution over bit strings $r$ of the proper length.[8]

Let $U_n$ denote a random variable uniformly distributed over $\{0,1\}^n$.

Suppose that our algorithm $A$ requires $m$ truly random bits to process a given input $x$. Furthermore, suppose that we can define a function $Ext: \{0,1\}^n \times \{0,1\}^d \to \{0,1\}^m$ such that if $X$ is our physical source and $U_d$ is uniformly distributed over $\{0,1\}^d$ then $Ext(X, U_d)$ is uniformly distributed over $\{0,1\}^m$. Here is a way to simulate $A()$ using $X$: (i) get a sample $x \sim X$, (ii) for every $s \in \{0,1\}^d$, compute $a_s := A(Ext(x,s), I)$, (iii) output the most common value among the $a_s$.

It is now easy to show that the above algorithm computes $f(I)$ with probability at least .8, over the choice of $X$. This is because $\mathbb{P}[A(Ext(U_d, X), I) = f(I)] \geq .9$ and so

$$\mathbb{P}_X \left[ \mathbb{P}_{U_d}[A(Ext(U_d, X), I) = f(I)] > \frac{1}{2} \right] \geq .8 \qquad (2)$$

---

[8]The reader may find .9 to be a poor formalization of the notion of "with *high* probability," but it is easy to reduce the error probability at the cost of a moderate increase of the running time.

The running time of our simulation of $A$ is $2^d$ times the running time of $A$, which is polynomial in the running time of $A$ provided that $d$ is logarithmic.

For this reasoning to work, it is not necessary that $Ext(X, U_d)$ be distributed *exactly* uniformly, but it is enough if it approximates the uniform distribution in an appropriate technical sense. If $X$ and $Y$ are two random variables taking values in $\Omega$, then we define their *variational distance* (also called *statistical distance* as

$$||X - Y||_{\text{SD}} := \max_{T \subseteq \Omega} |\mathbb{P}[X \in T] - \mathbb{P}[Y \in T]| \qquad (3)$$

We will sometimes call sets $T \subseteq \Omega$ *statistical tests*. If $||X - Y||_{\text{SD}} \leq \epsilon$ then we say that $X$ is $\epsilon$-close to $Y$.

We say that $Ext : \{0,1\}^n \to \{0,1\}^d \to \{0,1\}^m$ is a *seeded extractor* for a distribution $X$ with *error parameter* $\epsilon$ if $Ext(X, U_d)$ is $\epsilon$-close to $U_m$.

Vazirani and Vazirani [79] provided extractors for a certain class of distributions. (Their terminology was different.) Zuckerman [86] was the first to show that extractors exist for a very general class of distributions. Define the min-entropy of $X$ as $H_\infty(X) := \min_a \log_2 \frac{1}{\mathbb{P}[X=a]}$. If $H_\infty(X) \geq k$, then we say that $X$ is a $k$-source.

**Definition 3.1.** *A function $Ext : \{0,1\}^n \to \{0,1\}^d \to \{0,1\}^m$ is a $(k, \epsilon)$ seeded extractor if $Ext(X, U_d)$ is $\epsilon$-close to $U_m$ for every $k$-source $X$.*

The definition is implicit in [86]. The term *extractor* was coined in [45]. The term "seeded" refer to the truly random input of length $d$, which is called a *seed*. From now, we will refer to seeded extractors as simply "extractors."

Let $Ext : \{0,1\}^n \times \{0,1\}^d \to \{0,1\}^m$ be a $(k, \epsilon)$-extractor. Construct a bipartite graph $G = ([N], [M], E)$ with $N = 2^n$ vertices on the left, $M = 2^m$ vertices on the right. Connect two vertices $u, v$ if there is an $s$ that $v = Ext(u, s)$. Then if we pick any subset $S \subseteq [N]$ on the left and any subset $T \subseteq [M]$ on the right, the number of edges is $|S| \cdot 2^d \cdot |T|/2^m$ plus or minus $\epsilon|S|2^d$, provided $|S| \geq 2^k$. This is similar to one of the definitions of expander. Zuckerman and Wigderson [83] prove that one can derive expanders with very strong "edge expansion" from extractors.

Radakrishnan and Ta-Shma show that, in every extractor, $d \geq \log(n - k) + 2\log(1/\epsilon) - O(1)$ and that $m \leq k + d - O(1)$. Non-constructively, one can show that such bounds are achievable up to the additive constant factor, but explicit constructions are difficult. We will discuss explicit constructions later.

## 3.3 Applications

Randomness extractors have several applications, some of which are described below. See the tutorial by Salil Vadhan [76] and the survey by Ronen Shaltiel [57] for more examples and a broader discussion.

**Simulation of randomized algorithms.** Suppose we a randomized algorithm $A$ that on input $I$ computes $f(I)$ with probability, say, .9, and suppose that $Ext$ is a $(k', 1/4)$ extractor and that $X$ is a $k$-source. As before, let us sample $x \sim X$ and compute $A(Ext(x,s), I)$ for every $s$ and output the majority value. Let $B$ be the set of $x$ such that the algorithm fails. If $|B| \geq 2^{k'}$, then consider a random variable $Y$ uniformly distributed over $B$. It has entropy $k$, so $Ext(Y, U_d)$ should be $1/4$-close to uniform. Consider the statistical test $T$ defined as

$$T := \{r : A(r, I)\} = f(I)\} \tag{4}$$

Then $\mathbb{P}[U_n \in T] \geq .9$ by assumption and $\mathbb{P}[Ext(Y, U_d) \in T] \leq 1/2$ by construction. This would contradict $Ext$ being an extractor. We then conclude that $|B| \leq 2^{k'}$, and so that probability that our algorithm fails is at most $\mathbb{P}[X \in B] \leq |B|/2^k \leq 2^{k'-k}$.

This is very useful even in a setting in which we assume access to a perfect random source. In such a case, by using $n$ truly random bits we achieve an error probability that is only $2^{k'-n}$. Note that, in order to achieve the same error probability by running the algorithm several times independently we would have used $O((n-k') \cdot m)$ random bits instead of $n$.

**Other applications.** Randomness extractors are also very useful in settings where we assume a fully random distribution, say, over $n$ bits, that is *unknown* to us, except for some partial information whose entropy is at most $n - k$ bits. Then the distribution of the unknown string *conditioned on our knowldge* still has entropy at least $k$. If an extractor is applied to the unknown string, then the output of the extractor will be uniformly distributed even conditioned on our knowledge. In other words, our knowledge in useless in gaining any information about the output of the extractor.

This approach is used in the cryptographic settings of *privacy amplification* and *everlasting security* and in the design of pseudorandom generators for space-bounded algorithms. See [39, 77] and the references therein for the application to everlasting security and [45, 30, 51] for the application to pseudorandom generators.

## 4 Circuit Complexity

In order to discuss the computational approach to pseudorandomness we need to define a measure of efficiency for algorithms. We will informally talk about the "running time" of an algorithm on a given input without giving a specific definition. The reader can think of it as the number of elementary operations performed by an implementation of the algorithm on a computer. A more formal definition would be the number of steps in a Turing machine implementation of the algorithm. (See e.g. [64] for a definition of Turing machine.)

We say that a set $L \subseteq \{0,1\}^*$ is *decidable* in time $t(n)$ if there is an algorithm that on input $x \in \{0,1\}^n$ decides in time $\leq t(n)$ whether $x \in L$.

We are also interested in a more "concrete" measure of complexity, called *circuit complexity*. For integers $n$ and $i \leq n$, define the set $P_{i,n} := \{(a_1, \ldots, a_n) \in \{0,1\}^n : a_i = 1\}$. We say that a set $S \subseteq \{0,1\}^n$ has a *circuit of size* $K$ if there is a sequence of sets $S_1, \ldots, S_K$ such that: (i) $S_K = S$ and (ii) each $S_j$ is either a set $P_{i,n}$, or it is the complement of a set $S_h$, $h < j$, or it is the union $S_h \cup S_\ell$ of two sets, with $h, \ell < j$ or it is the intersection $S_h \cap S_\ell$ of two sets, with $h, k < j$. We say that a function $f : \{0,1\}^n \to \{0,1\}$ has a circuit of size $K$ if it is the characteristic function of a set that has a circuit of size $K$.

The *circuit complexity* of a set $S$ is the minimum $K$ such that $S$ has a circuit of size $K$. (Similarly for boolean functions.)

It is easy to see that there are subsets of $\{0,1\}^n$ whose circuit complexity is at least $c\frac{2^n}{n}$ for some constant $c > 0$: if a set has circuit complexity at most $K$, then it can described by using only $O(K \log K)$ bits, and so there are $2^{O(K \log K)}$ sets of circuit complexity at most $K$. If this number is less than $2^{2^n}$ then there exists a set of circuit complexity larger than $K$. Indeed, by the same argument, a random set has circuit complexity at least $c\frac{2^n}{n}$ with very high probability.

If $L \subseteq \{0,1\}^*$ is a set decidable in time $t(n)$, then for every $n$ there is a circuit of size $O((t(n))^2)$ for $L \cap \{0,1\}^n$. This implies that in order to prove lower bounds on the running time of algorithms for a given decision problem it is enough to prove lower bounds to the circuit complexity of finite fragments of it.[9]

So far, there has been very little success in proving circuit complexity lower bounds for "explicit sets," such as sets in NP. The strongest known lower bound is $5n$ [38, 32], and even an $n \log n$ lower bound is considered hopelessly out of reach of current techniques.

This is perhaps surprising given the simplicity of the definition of circuit complexity. The definition looks like a finite version of the definition of complexity for Borel sets, and one may hope that one could transfer techniques from topology to this setting. Sipser describes this idea in [62, 63], but, unfortunately, so far it has not led to any lower bound for general circuits.

Complexity theorists' failure to prove strong circuit lower bounds is partly explained by a famous paper by Razborov and Rudich [52]. They describe a general class of approaches to lower bounds, that they call "natural proofs." Razborov and Rudich show that all known methods to prove lower bounds for restricted classes of circuits yield natural proofs, but that (under certain complexity-theoretic assumptions) natural proofs cannot prove lower bounds for general circuits. The complexity theoretic assumption is itself about circuit lower bounds, and it is used to construct certain pseudorandom generators. The pseudorandom generators, in turn, imply the impossibility result. Somewhat inaccurately, the Razborov-Rudich result can be summarized as:

> *Circuit lower bounds are difficult to prove because they are true.*

## 5 Pseudorandom Generators and Their Application to Derandomization

Informally, a pseudorandom generator is an efficiently computable map $G : \{0,1\}^t \to \{0,1\}^m$, where $m$ is much bigger than $t$, such that, for a uniformly selected $x \in \{0,1\}^t$, the distribution $G(x)$ is pseudorandom, that is, it "looks like" the uniform distribution over $\{0,1\}^m$. We begin by describing how to formalize the notion of a distribution "looking like" the uniform distribution, and, more generally, the notion of two distributions "looking like" one other.

Recall thatr we use $U_n$ to denote a random variable that is uniformly distributed in $\{0,1\}^n$.

Ideally, we would like to say that $G()$ is a good pseudorandom generator if $G(U_t)$ and $U_m$ are close in statistical distance. Then, as we already discussed in Section 3, every application in which $m$ truly random bits are needed could be realized using the output of the generator (with a small increase in the probability of error). Unfortunately, this is too strong a definition: consider the statistical test $T$ defined to be the set of all possible outputs of $G$. Then $\mathbb{P}[G(U_t) \in T] = 1$ but $\mathbb{P}[U_m \in T] \leq 2^{t-m}$.

---

[9]The converse is not true: one can have undecidable sets of bounded circuit complexity.

The great idea that came from the work of Blum, Goldwasser, Micali and Yao in 1982 [22, 9, 84] was to modify the notion of statistical distance by considering only *efficiently computable* statistical tests.

**Definition 5.1 (Computational Indistinguishability).** *Two distributions $\mu_X$ and $\mu_Y$ over $\{0,1\}^m$ are $(K, \epsilon)$-indistinguishable if for every set $T \subseteq \{0,1\}^m$ of circuit complexity at most $K$,*

$$\left| \mathop{\mathbb{P}}_{x \sim \mu_X}[x \in T] - \mathop{\mathbb{P}}_{y \sim \mu_Y}[y \in T] \right| \leq \epsilon$$

**Definition 5.2 (Pseudorandomness).** *A distribution $\mu_X$ over $\{0,1\}^m$ is $(K, \epsilon)$-pseudorandom if it is $(K, \epsilon)$-indistinguishable from the uniform distribution. That is, for every $T \subseteq \{0,1\}^m$ of circuit complexity $\leq K$,*

$$\left| \mathop{\mathbb{P}}_{x \sim \mu_X}[x \in T] - \frac{|T|}{2^m} \right| \leq \epsilon$$

The following definition is due to Nisan and Wigderson [44].

**Definition 5.3 (Quick Pseudorandom Generator).** *Suppose that for every $n$ there is a $G_n : \{0,1\}^{t(n)} \to \{0,1\}^n$ that is $(n^2, 1/n)$-pseudorandom, and that there is an algorithm $G$ that, given $n, s$, computes $G_n(s)$ in time $2^{O(t(n))}$. Then $G$ is called a $t(n)$-quick pseudorandom generator.*

Suppose that an $O(\log n)$-quick pseudorandom generator (abbreviated logQPRG) exists, and suppose that $f$ is a function and $A$ is a polynomial time randomized algorithm that computes $f$ with probability at least $3/4$. We now describe a derandomization of algorithm $A$.

Let $I$ be an input, and let $m$ be the number of random bits used by $A$ on input $I$. Let $K$ be an efficiently computable upper bound to the circuit complexity pf $T := \{r : A(r, I) = f(I)\}$. Choose $n$ to be large enough so that: (i) $n^2 \geq K$, (ii) $n \geq m$, and (iii) $n \geq 5$. Because of our assumption that $A$ runs in polynomial time, $n$ is polynomial in the length of $I$.[10]

Now, compute $A(G_n(s), I)$ for each $s$, and output the value that is returned most often. This completes the description of a polynomial time deterministic algorithms

Regarding correctness, we assumed $\mathbb{P}[A(U_m, I) = f(I)] \geq \frac{3}{4}$, and so

$$\mathbb{P}[A(G_n(U_{t(n)}), I) = f(I)] \geq \frac{3}{4} - \frac{1}{n} > \frac{1}{2} \tag{5}$$

Otherwise, the set $T = \{r : A(r, I) = f(I)\}$ contradicts the pseudorandomness of $G_n$. Something similar can be done if $A$ is only guaranteed to *approximate* $f$ with high probability, for example if $f(I)$ is the number of perfect matchings in the graph represented by $I$ and $A$ is the Jerrum-Sinclair-Vigoda probabilistic approximation algorithm for this problem [33]. The only difference is that we take the *median* of the outputs instead of the most common one.

The applications of logQPRGs to the probabilistic method is as follows. Suppose that:

---

[10]It should be noted that we may not know how to construct a circuit for $T$, because it seems that to construct such a circuit we need to know $f(I)$. In order to compute a polynomially bounded upper bounds to the circuit complexity of $T$, however, we just need to find out how large is the circuit for $T$ that *we would be able to build if we knew $f(I)$*.

- For every $n$, we have a set $\Omega_n$ of "objects of size $n$" (for example, graphs with $n$ vertices and maximum degree $d$, where $d$ is a fixed constant). It is convenient to assume the sets $\Omega_n$ to be disjoint.

- We define $P \subseteq \bigcup_n \Omega_n$ to be the set of *interesting* objects that we would like to construct. (For example, expander graphs.)

- Property $P$ is computable in polynomial time. That is, there is an algorithm that given $n$ and $x \in \Omega_n$ runs in time polynomial in $n$ and determines whether $x \in P$.

- The probabilistic method proves that such graphs exist and are "abundant." That is, for every $n$, we define a probability distribution $\mu_n$ over $\Omega_n$ and we prove that $\mathbb{P}_{x \sim \Omega_n}[x \in P] \geq \frac{1}{2}$. (The constant $1/2$ is not important.)

- The distributions $\mu_n$ are polynomial time samplable. That is, there is a probabilistic algorithm $A$ that, given $n$, generates in time polynomial in $n$ a sample from $\mu_n$.

This formalization captures the way the probabilistic method is typically used in practice, with the exception of the efficient computability of $P$, which sometimes is not true. (For example, in the problem of finding lower bounds for $R(k,k)$.) Finally, suppose that a logQPRG exists. Given $n$ here is how we construct an element in $P \cap \Omega_n$. Let $m$ be the number of random bits used by $A$ to sample an element of $\mu_n$, and let $K$ be an upper bound to the size of a circuit for the set $T := \{r : A(n,r) \in P\}$. As before, we can use the assumption that $A$ is computable in polynomial time and $P$ is decidable in polynomial time to conclude that $m$ and $K$ are upper bounded by polynomials in $n$. Let $N$ be large enough so that (i) $N \geq 3$, (ii) $N^2 \geq K$ and (iii) $N \geq m$. Then compute $A(n, G_N(N,s))$ for every $s$, and let $s_0$ be such that $A(n, G_N(N, s_0)) \in P$. Such an $s_0$ must exist, otherwise $T$ contradicts the pseudorandomness of $G_N$. Output $A(n, G_N(N, s_0))$.

## 6 Conditional Constructions of Pseudorandom Generators

Blum and Micali [9] construct $n^{o(1)}$QPRGs, according to a slightly different definition, assuming a specific number-theoretic assumption. Yao [84] proves that the Blum-Micali definition is equivalent to a definition based on indistinguishability and constructs $n^{o(1)}$QPRGs under the more general assumption that *one-way permutations* exist. Yao [84] also recognizes that $n^{o(1)}$QPRGs imply a $2^{n^{o(1)}}$ derandomization of every probabilistic algorithm.

Blum, Micali and Yao do not use the parametrization that we adopted in the definition of quick pseudorandom generators. In the cryptographic applications that motivate their work, it is important that the generator be computable in time polynomial in the length of the output (rather than exponential in the length of the input), and, if $m$ is the length of the output, one desires $(S(m), \epsilon(m))$-pseudorandomness where $S(m)$ and $1/\epsilon(m)$ are super-polynomial in $m$. Their constructions satisfy these stronger requirements.

Håstad et al. [26] show that the weaker assumption that *one-way functions* exist suffices to construct $n^{o(1)}$QPRGs. Their construction satisfies the stronger requirements of [9, 84]. We do not define one-way permutations and one-way functions here and we refer the interested reader to Goldreich's monograph [21], the definitive treatment of these results.

Nisan and Wigderson [44] introduced the definition of quick pseudorandom generator that we gave in the previous section and presented a new construction that works under considerably weaker assumptions than the existence of one-way functions.[11] The Nisan-Wigderson construction also "scales" very well, and it gives more efficient QPRGs if one is willing to start from stronger assumptions. A sufficiently strong assumption implies optimal logQPRGs, and this is the only version of the Nisan-Wigderson results that we will discuss.

We first need to define a notion of *average-case circuit complexity*. We say that a set $S \subseteq \{0,1\}^n$ is $(K,\epsilon)$-hard on average if for every set $T$ computable by a circuit of size $\leq K$ we have $\mathbb{P}[1_S(x) = 1_T(x)] \leq \frac{1}{2} + \epsilon$, where we use the notation $1_S$ for the characteristic function of the set $S$. We say that a set $L \subseteq \{0,1\}^*$ is $(K(n), \epsilon(n))$-hard on average if, for every $n$, $L \cap \{0,1\}^n$ is $(K(n), \epsilon(n))$-hard on average.

**Theorem 6.1 (Nisan and Wigderson [44]).** *Suppose there is a set $L$ such that: (i) $L$ can be decided in time $2^{O(n)}$ and (ii) there is a constant $\delta > 0$ such that $L$ is $(2^{\delta n}, 2^{-\delta n})$-hard on average. Then a $\log QPRG$ exists.*

When Theorem 6.1 was announced in 1988, average-case complexity was much less understood than worst-case complexity and it was not even clear if the assumption used in the Theorem was plausible.

This motivated a long-term research program on average-case complexity. Building on work by Babai, Fortnow, Impagliazzo, Nisan and Wigderson [6, 27], Impagliazzo and Wigderson finally proved in 1997 that the assumption of Theorem 6.1 is equivalent to a seemingly weaker *worst-case* assumption.

**Theorem 6.2 (Impagliazzo and Wigderson [31]).** *Suppose there is a set $L$ such that: (i) $L$ can be decided in time $2^{O(n)}$ and (ii) there is a constant $\delta > 0$ such that the circuit complexity of $L$ is at least $2^{\delta n}$.*

*Then there is a set $L'$ such that: (i) $L'$ can be decided in time $2^{O(n)}$ and (ii) there is a constant $\delta' > 0$ such that $L'$ is $(2^{\delta' n}, 2^{-\delta' n})$-hard on average.*

In conclusion, we have optimal logQPRG, and polynomial time derandomization of probabilistic algorithms, under the assumptions that there are problems of exponential circuit complexity that are computable in exponential time. Such an assumption is considered very plausible.

There are other applications of these techniques that we will not have space to discuss, including extensions to the case of pseudorandomness against "non-deterministic statistical tests," which imply surprising results for the Graph Isomorphism problem [35, 43].

## 7 Average-Case Complexity and Codes

We now come to a connection between the Impagliazzo-Wigderson Theorem and error-correcting codes. Due to space limitations we will only give a short discussion. The interested reader is referred to our survey paper [72] for more details.

---

[11] On the other hand, the Nisan-Wigderson generator does not satisfy the stronger properties of the pseudorandom generators of Blum, Micali, Yao, Håstad et al. [9, 84, 26]. This is unavoidable because the existence of such stronger pseudorandom generators is *equivalent* to the existence of one-way functions.

Impagliazzo and Wigderson derive Theorem 6.2 from the following "hardness amplification" reduction.

**Theorem 7.1 (Impagliazzo and Wigderson [31]).** *For every $\delta > 0$ there are constants $\delta' > 0$, $c > 1$, and an algorithm with the following property.*

*If $S \subseteq \{0,1\}^n$ is a set of circuit complexity at least $2^{\delta n}$, then, on input $S$, the algorithm outputs a set $S' \subseteq \{0,1\}^{cn}$ that is $(2^{\delta' n}, 2^{-\delta' n})$ hard on average.*

Like most results in complexity theory, the proof is by contradiction: suppose we have a set $T$ computable by a circuit of size $2^{\delta' n}$ such that $\mathbb{P}_{x \sim \{0,1\}^n}[1_{S'}(x) = 1_T(x)] \geq 1/2 + 2^{-\delta' n}$; then Impagliazzo and Wigderson show how to use such the circuit for $T$ to construct a circuit for $S$ of size $2^{\delta n}$.

Phrased this way, the result has a strong coding-theoretic flavor: we can think of $S$ as a "message," of $S'$ as the "encoding" of $S$, of $T$ as the "corrupted transmission" that the receiver gets, and of the process of reconstructing (a circuit for) $S$ from (a circuit for) $T$ as a "decoding" process. Given this perspective, introduced in [68], it is natural to try and apply coding-theoretic algorithms to hardness amplification. In doing so, we encounter the following difficulty: viewed as a message, a set $S \subseteq \{0,1\}^n$ is (or can be represented as) a bit-string of length $N = 2^n$, and so a polynomial time coding-theoretic algorithm that reconstructs $S$ from a corrupted encoding of $S$ takes time $N^{O(1)} = 2^{O(n)}$. In Theorem 7.1, however we need to produce a circuit of size $2^{\delta n} = N^{\delta}$, and so the circuit cannot simply be an implementation of the decoding algorithm.

It seems that what we need the following type of error-correcting code (we use the notation $\mathcal{P}(A)$ to denote the set of all subsets of a set $A$): a map $C : \mathcal{P}(\{0,1\}^n) \to \mathcal{P}(\{0,1\}^{n'})$, with $n' = O(n)$ such that there is an algorithm that given a set $T \in \mathcal{P}(\{0,1\}^{n'})$ close to the encoding $C(S)$ of a message $S \in \mathcal{P}(\{0,1\}^{n'})$ and an elements $a \in \{0,1\}^n$ determines in time at most $2^{\delta n}$ whether $a \in S$ or not. If we think of a set $S \in \mathcal{P}(\{0,1\}^n)$ as simply a bit-string in $\{0,1\}^N$, $N = 2^n$, then we are looking for an error correcting code $C : \{0,1\}^N \to \{0,1\}^{N'}$, with $N' = N^{O(1)}$, such that there is an algorithm that given a string $u \in \{0,1\}^{N'}$ close to an encoding $C(x)$, and given an index $i \in \{1, \ldots, N\}$, computes in time at most $N^{\delta}$ the bit $x_i$. It remains to specify how to "give in input" a string $u$ of length $N' > N$ to an algorithm of running time, say, $N^{.001}$: the algorithm does not even have enough time to *read* the input. This can be handled by modeling the input as an "oracle" for the algorithm, which is a standard notion.

The existence of error-correcting codes with this kind of "sub-linear time decoding algorithms" was well known, but the problem is that this notion is still not sufficient for the application to Theorem 7.1. The reason is that we have described a decoding algorithm that gives a unique answer and, as discussed in Section 2, such algorithms cannot recover from more than a $1/4 + o(1)$ fraction of errors. Theorem 7.1, however, requires us to correct from close to $1/2$ fraction of errors.

In Section 2 we remarked that it is possible to do *list*-decoding even after almost a $1/2$ fraction of errors occur. So we need a definition of *sub-linear time list decoding algorithm*. The definition is too technical to give here. It was formulated, for a different application, in [5]. A reasonably simple sub-linear time list-decoding algorithm tgiving a new proof Theorem 7.1 is presented in [68]. The coding-theoretic proof is considerably simpler than the original one.

The connection between error-correcting and hardness amplification also goes in the other direction: it is possible to view the techniques of [6, 27, 31] as defining list-decodable codes with sub-linear time decoding algorithm. This reverse connection has been used to transfer known known coding theoretic impossibility results to the setting of amplification of hardness.

Recall that if we want to correct from $1/2 - \epsilon$ errors, then unique decoding is impossible. Codes that are $(L, 1/2 - \epsilon)$-list decodable exist, but it is possible to prove that for such codes we need $L = \Omega(\epsilon^{-2})$. In our proof [68] of Theorem 6.2, this is not a problem because when we realize the decoding algorithm as a circuit we can "hard-wire" into the circuit the correct choice of element from the list. Suppose, however, that we want to prove a version of Theorem 6.2 where "algorithm of running time $K$" replaces "circuits of size $K$." Then such a theorem would not follow from [68]: if we try to follow the proof we see that from a good-on-average algorithm for $L' \cap \{0,1\}^{n'}$ we can only construct a *list of algorithms such that one of them* computes $L \cap \{0,1\}^n$ correctly, and it is not clear how to choose one algorithm out of this list.[12] This problem is solved in [74], where we do prove a version of Theorem 6.2 with "probabilistic algorithm" in place of "circuit."

Viola [81] proves that error-correcting codes cannot be computed in certain very low complexity classes, and this means that the exponentially big error-correcting code computations occurring in [68] must add a very strong complexity overhead. This means that coding-theoretic techniques cannot prove a version of Theorem 6.2 where "computable in time $2^{O(n)}$" is replaced by "computable in NP." Indeed, it remains a fundamental open question whether a theorem showing equivalence of worst-case complexity and average-case complexity in NP can be proved. Results of [19, 10] show that this is unlikely.

Impagliazzo [28] wonders about a *positive* use of the fact that amplification of hardness results imply error-correcting codes, and whether the techniques of [6, 27, 31] would lead to practical error-correcting codes. We explore this question in [71], focusing on an optimization of the techniques of [27], but our results are far from being competitive with knwon constructions and algorithms of list-decodable codes. On the other hand, our work in refining the techniques of [27], while not successful in deriving good coding-theoretic applications, has led to interesting applications within complexity theory [71, 73].

## 8 Extractors and Pseudorandom Generators

We now come to what is perhaps the most surprising result of this survey, the fact that (the proofs of) Theorems 6.1 and 6.2 directly lead to *unconditional* constructions of extractors.

First, let us give a very high-level description of the pseudorandom generator construction that follows from Theorems 6.1 and 6.2.

Let $L$ be the set of exponential circuit complexity as in the assumption of Theorem 6.2, and let $m$ be a parameter such that we are looking to construct a generator $G_m : \{0,1\}^{O(\log m)} \to \{0,1\}^m$ whose output is $(m^2, 1/m)$ pseudorandom. First, we define $\ell = O(\log m)$ such that $L \cap \{0,1\}^\ell$ has circuit complexity at least $m^c$, for a certain absolute constant $c$. Then we define our generator as $G_m(z) = IW_m(L \cap \{0,1\}^\ell, z)$, where $IW_m(S, z)$ is a procedure that takes in input a set $S \subseteq \{0,1\}^\ell$ and a string $z \in \{0,1\}^{O(\log m)}$, output a string in $\{0,1\}^m$, and it is such that if $S$ has circuit complexity at least $m^c$ then $IW_m(S, U_{O(\log m)})$ is $(m^2, 1/m)$ pseudorandom. Proving that $IW_m(\cdot, \cdot)$ has this property is of course quite complicated, but the general outline is as follows. As usual we proceed by contradiction, and start from a statistical test $T$ of circuit complexity at most $m^2$ such that, supposedly,

$$|\mathbb{P}[U_m \in T] - \mathbb{P}[IW_m(S, U_{O(\log m)}) \in T]| > \frac{1}{m}$$

---

[12]This difficulty is discussed in [74].

Then we modify the circuit for $T$ and build a new circuit for $S$ of size $< m^c$, thus contradicting the hypothesis.

The analysis, indeed, proves a more general result. We will need some additional definitions before stating this more general result. For sets $T \subseteq \{0,1\}^m$ and $S \subseteq \{0,1\}^\ell$, we say that $S$ has a *circuit with $T$-gates of size $K$* if there is a sequence of sets $S_1, \ldots, S_m$ such that $S_m = S$, and each $S_j$ is either a set of the form $P_{i,n}$, or it is the complement of a set $S_h$ $h < j$, or it is the union or the intersection of two sets $S_h, S_{h'}$ with $h, h' < j$, or it is defined as

$$S_j := \{a \in \{0,1\}^\ell : (1_{S_{h_1}}(a), \ldots, 1_{S_{h_m}}(a)) \in T\}$$

for some $h_1, \ldots, h_m < j$. It is not hard to show that if $S$ has a circuit with $T$-gates of size $K_1$, and $T$ has a regular circuit of size $K_2$, then $S$ has a regular circuit of size at most $K_1 \cdot K_2$. With these definitions in place we can be more specific about the analysis in [44, 31]: the analysis shows that if $S \subseteq \{0,1\}^\ell$ and $T \subseteq \{0,1\}^m$ are two *arbitrary* sets such that

$$|\mathbb{P}[U_m \in T] - \mathbb{P}[IW_m(S, U_{O(\log m)}) \in T]| > \frac{1}{m}$$

then there is a circuit with $T$-gates for $S$ of size $< m^{c-2}$. (Note that this implies our previous statement.)

Here is the main idea in [70]: suppose that we have access to a weak random source, that is, a random variable $X$ taking values in $\{0,1\}^n$ and having min-entropy at least $k$. Suppose that $n = 2^\ell$. Then we can, equivalently, see $X$ as being distributed over $\mathcal{P}(\{0,1\}^\ell)$, the set of all subsets of $\{0,1\}^\ell$. What can we say about the distribution of $IW_m(X, U_{O(\log m)})$? We claim that, if $k$ is large enough, the distribution $IW_m(X, U_{O(\log m)})$ is close in statistical distance to the uniform distribution; in other words, $IW_m(\cdot, \cdot)$ is an *extractor*.

Let us see how to prove this by contradiction. Let $T$ be a statistical test such that

$$|\mathbb{P}[U_m \in T] - \mathbb{P}[IW_m(X, U_{O(\log m)}) \in T]| > \frac{1}{m}$$

and call a set $S \in \mathcal{P}(\{0,1\}^\ell)$ *bad* if

$$|\mathbb{P}[U_m \in T] - \mathbb{P}[IW_m(S, U_{O(\log m)}) \in T]| > \frac{2}{m}$$

Let $B$ be the set of all bad sets. Then, by Markov's inequality,

$$\mathbb{P}[X \in B] \geq \frac{2}{m}$$

and since $X$ has min-entropy $k$ we have $|B| \geq 2^{k - \log m - 1}$. On the other hand, if $S$ is bad, then there is a circuit with $T$-gates of size at most $m^{c-2}$ that computes $S$. The number of such circuits is at most $2^{O(m^{c-1})}$, and so $|B| \leq 2^{O(m^{c-1})}$. So if $k \geq c'm^{c-1}$, where $c, c'$ are absolute constants, we reach a contradiction. Thus,

$$||IW_m(X, U_{O(\log m)}) - U_m||_{\text{SD}} \leq \frac{1}{m}$$

If we look more closely at how $IW_m(S, z)$ is defined, we see that (especially if we use the [68] proof of Theorem 6.2) it can be seen as $IW_m(S, z) := NW_m(C(S), z)$, where $C()$ is an error-correcting code and $NW_m$ is the relatively simple pseudorandom generator construction of Nisan and Wigderson. For the application to derandomization, it is important that $C()$ be a "sub-linear time list-decodable" error-correcting code. In order for our argument about randomness extraction to work, however, it is sufficient that $C()$ be an arbitrary list-decodable code, and not even a polynomial time list-decoding algorithm is needed. This means that one can get extractors by using standard error-correcting codes and the simple Nisan-Wigderson generator. The resulting construction is described and analysed in [70] in about two pages and, at the time, it was the best known extractor construction, improving over very technical previous work.

What makes these calculations work is the intuition that the proofs of Theorems 6.1 and 6.2 prove more than the intended statement. In particular, the proof works if we replace "circuit complexity" with "description complexity" which what we exploited in the previous argument. See [70] for further discussion of this point.

The connection with pseudorandomness and the general idea of analysing an extractor by finding short descriptions of the output of the source based on a hyptothetical statistical test (the so-called "reconstruction method" to analyse extractors) has led to remarkable advances in extractor constructions in the past five years, together with other ideas. The best distillation of the reconstruction method is in [58], providing a near-optimal and simple construction of extractors.[13] The extractor motivation has also led to improvements in pseudorandom generator constructions, see [58, 75]. Currently, the best known extractor construction [40] uses the notion of "condenser" introduced in [69, 54] and a combination of several components, one of which is analysed with the reconstruction method. The extractors of Lu et al. [40] is almost best possible.

## 9 Conclusions

We have discussed how, starting from worst-case complexity assumptions, it is possible to construct very strong pseudorandom generators, and derive conditional derandomization results for *all* probabilistic algorithms.

What about *proving* circuit lower bounds and deriving unconditional derandomization results? The results of Razborov and Rudich [52] show that a significant departure from current techniques will be required to prove such lower bounds. What about deriving derandomization results *without* proving lower bounds? Impagliazzo, Kabanets and Wigderson [29] prove that any general derandomization result implies circuit lower bound.[14]

---

[13] The construction is simple but the analysis is quite non-trivial.

[14] Here is what we mean by "general derandomization:" if $f$ is a function and $A$ is randomized algorithm that with high probability achieves a good approximation of $f$, then there is a deterministic algorithm that achieves a good approximation of $f$ and whose running time is polynomial in the running time of $A$.

Short of proving such elusive circuit lower bounds, we should test the prediction of the theory and look for polynomial time deterministic versions of known probabilistic polynomial time algorithms. The four most important probabilistic algorithms (or collections of algorithms) are: primality testing, graph connectivity using random walks, polynomial identity testing, and algorithms for approximate counting. Primality testing and graph connectivity using random walks have been derandomized [1, 53]. Kabanets and Impagliazzo [34] prove that any derandomized polynomial identity testing algorithms implies circuit lower bounds.[15]

The possibility of derandomizing approximate counting algorithms with current techniques is quite open. Here is perhaps the simplest question: given an $n$-variable boolean formula in disjunctive normal form and $\epsilon > 0$, compute in time polynomial in the size of the formula and in $1/\epsilon$ an approximation to the number of satisfying assignments up to an additive error $\leq 2^n \epsilon$. See [42] for a nearly polynomial time deterministic algorithm for this problem.

The construction of an optimal (seeded) extractor with parameters matching the known lower bounds remains an elusive open question. It would also be interesting to match the parameters of [40] with a simpler construction.

There has been very exciting recent progress towards constructing good seedless extractors for independent sources, and for the related problem of constructing bipartite Ramsey graphs [8, 11]. The broader area of seedless extractor constructions for general classes of distributions has seen much recent progress. In the long run, we would expect this research to define simple and powerful seedless extractors working for a wide and natural class of distributions. Such extractors would be very useful in practice, giving a principled approach to the production of random bits for cryptographic applications.

# References

[1] Manindra Agrawal, Neeraj Kayal, and Nitin Saxena. PRIMES is in P. *Annals of Mathematics*, 160(2):781–793, 2004. 4, 19

[2] Romas Aleliunas, Richard M. Karp, Richard J. Lipton, László Lovász, and Charles Rackoff. Random walks, universal traversal sequences, and the complexity of maze problems. In *Proceedings of the 20th IEEE Symposium on Foundations of Computer Science*, pages 218–223, 1979. 2

[3] Noga Alon and Michael R. Capalbo. Explicit unique-neighbor expanders. In *Proceedings of the 43rd IEEE Symposium on Foundations of Computer Science*, pages 73–79, 2002. 7

[4] Noga Alon and Joel Spencer. *The Probabilistic Method*. John Wiley and Sons, 2000. 2

[5] Sanjeev Arora and Madhu Sudan. Improved low degree testing and its applications. *Combinatorica*, 23(3):365–426, 2003. 15

---

[15] Fortunately, these are not of the kind ruled out by [52], so there is some hope. Indeed Raz [49, 50] has recently proved lower bounds that are weaker than, but in the spirit of, what is needed to derandomize polynomial identity testing.

[6] László Babai, Lance Fortnow, Noam Nisan, and Avi Wigderson. BPP has subexponential time simulations unless EXPTIME has publishable proofs. *Computational Complexity*, 3(4):307–318, 1993. 14, 15, 16

[7] Boaz Barak, Russell Impagliazzo, and Avi Wigderson. Extracting randomness using few independent sources. In *Proceedings of the 45th IEEE Symposium on Foundations of Computer Science*, pages 384–393, 2004. 8

[8] Boaz Barak, Guy Kindler, Ronen Shaltiel, Benny Sudakov, and Avi Wigderson. Simulating independence: new constructions of condensers, Ramsey graphs, dispersers, and extractors. In *Proceedings of the 37th ACM Symposium on Theory of Computing*, pages 1–10, 2005. 8, 19

[9] Manuel Blum and Silvio Micali. How to generate cryptographically strong sequences of pseudorandom bits. *SIAM Journal on Computing*, 13(4):850–864, 1984. Preliminary version in *Proc. of FOCS'82*. 3, 12, 13, 14

[10] Andrej Bogdanov and Luca Trevisan. On wost-case to average-case reductions for NP problems. In *Proceedings of the 44th IEEE Symposium on Foundations of Computer Science*, pages 308–317, 2003. 16

[11] Jean Bourgain. More on the sum-product phenomenon in prime fields and its applications. *International Journal of Number Theory*, 1(1):1–32, 2005. 8, 19

[12] Jean Bourgain, Nets Katz, and Terence Tao. A sum-product estimate for finite fields, and applications. *Geometric and Functional Analysis*, 14:27–57, 2004. 8

[13] Michael R. Capalbo, Omer Reingold, Salil P. Vadhan, and Avi Wigderson. Randomness conductors and constant-degree lossless expanders. In *Proceedings of the 34th ACM Symposium on Theory of Computing*, pages 659–668, 2002. 7

[14] Benny Chor and Oded Goldreich. Unbiased bits from sources of weak randomness and probabilistic communication complexity. *SIAM Journal on Computing*, 17(2):230–261, April 1988. 8

[15] Fan R. K. Chung, Ronald L. Graham, and Richard M. Wilson. Quasi-random graphs. *Combinatorica*, 9(4):345–362, 1989. 7

[16] Aviad Cohen and Avi Wigderson. Dispersers, deterministic amplification, and weak random sources. In *Proceedings of the 30th IEEE Symposium on Foundations of Computer Science*, pages 14–19, 1989. 8

[17] Peter Elias. List decoding for noisy channels. Technical Report 335, Research Laboratory of Electronics, MIT, 1957. 6

[18] Paul Erdős. Some remarks on the theory of graphs. *Bulletin of the AMS*, 53:292–294, 1947. 1

[19] Joan Feigenbaum and Lance Fortnow. Random-self-reducibility of complete sets. *SIAM Journal on Computing*, 22:994–1005, 1993. 16

[20] Peter Frankl and Richard M. Wilson. Intersection theorems with geometric consequences. *Combinatorica*, 1(4):357–368, 1981. 2

[21] Oded Goldreich. *The Foundations of Cryptography - Volume 1*. Cambridge University Press, 2001. 13

[22] Shafi Goldwasser and Silvio Micali. Probabilistic encryption. *Journal of Computer and System Sciences*, 28(2):270–299, 1984. Preliminary Version in *Proc. of STOC'82*. 3, 12

[23] Venkatesan Guruswami. *List Decoding of Error-Correcting Codes*. PhD thesis, MIT, 2001. 6

[24] Venkatesan Guruswami and Atri Rudra. Explicit capacity-achieving list-decodable codes. Technical Report TR05-133, Electronic Colloquium on Computational Complexity, 2005. 6

[25] Richard Hamming. Error detecting and error correcting codes. *Bell System Technical Journal*, 29:147–160, 1950. 6

[26] Johan Håstad, Russell Impagliazzo, Leonid Levin, and Michael Luby. A pseudorandom generator from any one-way function. *SIAM Journal on Computing*, 28(4):1364–1396, 1999. 13, 14

[27] Russell Impagliazzo. Hard-core distributions for somewhat hard problems. In *Proceedings of the 36th IEEE Symposium on Foundations of Computer Science*, pages 538–545, 1995. 14, 15, 16

[28] Russell Impagliazzo. Hardness as randomness: a survey of universal derandomization. In *Proceedings of the International Congress of Mathematicians*, volume 3, pages 659–672, 2002. 4, 16

[29] Russell Impagliazzo, Valentine Kabanets, and Avi Wigderson. In search of an easy witness: exponential time vs. probabilistic polynomial time. *Journal of Computer and System Sciences*, 65(4):672–694, 2002. 18

[30] Russell Impagliazzo, Noam Nisan, and Avi Wigderson. Pseudorandomness for network algorithms. In *Proceedings of the 26th ACM Symposium on Theory of Computing*, pages 356–364, 1994. 10

[31] Russell Impagliazzo and Avi Wigderson. $P = BPP$ unless $E$ has sub-exponential circuits. In *Proceedings of the 29th ACM Symposium on Theory of Computing*, pages 220–229, 1997. 3, 14, 15, 16, 17

[32] Kazuo Iwama and Hiroki Morizumi. An explicit lower bound of $5n - o(n)$ for boolean circuits. In *Proceedings of the 27th Symposium on Mathematical Foundations of Computer Science*, pages 353–364, 2002. 11

[33] Mark Jerrum, Alistair Sinclair, and Eric Vigoda. A polynomial-time approximation algorithm for the permanent of a matrix with nonnegative entries. *Journal of the ACM*, 51(4):671–697, 2004. 12

[34] Valentine Kabanets and Russell Impagliazzo. Derandomizing polynomial identity tests means proving circuit lower bounds. *Computational Complexity*, 13(1-2):1–46, 2004. 19

[35] Adam Klivans and Dieter van Melkebeek. Graph nonisomorphism has subexponential size proofs unless the polynomial-time hierarchy collapses. *SIAM Journal on Computing*, 31(5):1501–1526, 2002. 14

[36] Sergei Konyagin. A sum-product estimate in fields of prime order. arXiv:math.NT/0304217, 2003. 8

[37] Michael Krivelevich and Benny Sudakov. Pseudo-random graphs. Preprint, 2005. 7

[38] Oded Lachish and Ran Raz. Explicit lower bound of $4.5n - o(n)$ for boolean circuits. In *Proceedings of the 33rd ACM Symposium on Theory of Computing*, pages 399–408, 2001. 11

[39] Chi-Jen Lu. Encryption against storage-bounded adversaries from on-line strong extractors. *Journal of Cryptology*, 17(1):27–42, 2004. 10

[40] Chi-Jen Lu, Omer Reingold, Salil P. Vadhana, and Avi Wigderson. Extractors: optimal up to constant factors. In *Proceedings of the 35th ACM Symposium on Theory of Computing*, pages 602–611, 2003. 18, 19

[41] Alexander Lubotzky, R. Phillips, and Peter Sarnak. Ramanujan graphs. *Combinatorica*, 8:261–277, 1988. 7

[42] Michael Luby and Boban Velickovic. On deterministic approximation of DNF. *Algorithmica*, 16(4/5):415–433, 1996. 19

[43] Peter B. Miltersen and N.V. Vinodchandran. Derandomizing Arthur-Merlin games using hitting sets. In *Proceedings of the 40th IEEE Symposium on Foundations of Computer Science*, pages 71–80, 1999. 14

[44] Noam Nisan and Avi Wigderson. Hardness vs randomness. *Journal of Computer and System Sciences*, 49:149–167, 1994. Preliminary version in *Proc. of FOCS'88*. 3, 12, 14, 17

[45] Noam Nisan and David Zuckerman. Randomness is linear in space. *Journal of Computer and System Sciences*, 52(1):43–52, 1996. Preliminary version in *Proc. of STOC'93*. 4, 9, 10

[46] Farzad Parvaresh and Alexander Vardy. Correcting errors beyond the Guruswami-Sudan radius in polynomial time. In *Proceedings of the 46th IEEE Symposium on Foundations of Computer Science*, pages 285–294, 2005. 6

[47] Michael Rabin. Probabilistic algorithm for testing primality. *Journal of Number Theory*, 12:128–138, 1980. 2

[48] Anup Rao. Extractors for a constant number of polynomial min-entropy independent sources. Technical Report TR05-106, Electronic Colloquium on Computational Complexity, 2005. 8

[49] Ran Raz. Multi-linear formulas for permanent and determinant are of super-polynomial size. In *Proceedings of the 36th ACM Symposium on Theory of Computing*, pages 633–641, 2004. 19

[50] Ran Raz. Multilinear-$NC_1 \neq$ multilinear-$NC_2$. In *Proceedings of the 45th IEEE Symposium on Foundations of Computer Science*, pages 344–351, 2004. 19

[51] Ran Raz and Omer Reingold. On recycling randomness in space bounded computation. In *Proceedings of the 31st ACM Symposium on Theory of Computing*, pages 159–168, 1999.  10

[52] Alexander A. Razborov and Steven Rudich. Natural proofs. *Journal of Computer and System Sciences*, 55(1):24–35, 1997.  11, 18, 19

[53] Omer Reingold. Undirected ST-connectivity in log-space. In *Proceedings of the 37th ACM Symposium on Theory of Computing*, pages 376–385, 2005.  4, 8, 19

[54] Omer Reingold, Ronen Shaltiel, and Avi Wigderson. Extracting randomness by repeated condensing. In *Proceedings of the 41st IEEE Symposium on Foundations of Computer Science*, 2000.  18

[55] Miklos Santha and Umesh Vazirani. Generating quasi-random sequences from slightly random sources. *Journal of Computer and System Sciences*, 33:75–87, 1986.  8

[56] Jacob T. Schwartz. Fast probabilistic algorithms for verification of polynomial identities. *Journal of the ACM*, 27:701–717, 1980.  2

[57] Ronen Shaltiel. Recent developments in extractors. *Bulletin of the European Association for Theoretical Computer Science*, 77:67–95, 2002.  9

[58] Ronen Shaltiel and Christopher Umans. Simple extractors for all min-entropies and a new pseudorandom generator. *Journal of the ACM*, 52(2):172–216, 2005.  18

[59] Claude Shannon. A mathematical theory of communications. *Bell System Technical Journal*, 27:379–423, 623–656, 1948.  2

[60] Claude Shannon. The synthesis of two-terminal switching circuits. *Bell System Technical Journal*, 28:59–98, 1949.  2

[61] Alistair Sinclair and Mark Jerrum. Approximate counting, uniform generation and rapidly mixing Markov chains. *Information and Computation*, 82(1):93–133, 1989.  2

[62] Michael Sipser. Borel sets and circuit complexity. In *Proceedings of the 15th ACM Symposium on Theory of Computing*, pages 61–69, 1983.  11

[63] Michael Sipser. A topological view of some problems in complexity theory. In *Proceedings of the Symposium on Mathematical Foundations of Computer Science*, pages 567–572, 1984.  11

[64] Michael Sipser. *Introduction to the Theory of Computation*. PWS, 1997.  10

[65] Robert Solovay and Volker Strassen. A fast Monte-Carlo test for primality. *SIAM Journal on Computing*, 6(1):84–85, 1977.  2

[66] Madhu Sudan. Decoding of Reed-Solomon codes beyond the error-correction bound. *Journal of Complexity*, 13(1):180–193, 1997.  6

[67] Madhu Sudan. List decoding: Algorithms and applications. *SIGACT News*, 31(1):16–27, 2000.  6

[68] Madhu Sudan, Luca Trevisan, and Salil Vadhan. Pseudorandom generators without the XOR lemma. *Journal of Computer and System Sciences*, 62(2):236–266, 2001.   15, 16, 18

[69] Amnon Ta-Shma, Christopher Umans, and David Zuckerman. Loss-less condensers, unbalanced expanders, and extractors. In *Proceedings of the 33rd ACM Symposium on Theory of Computing*, 2001.   18

[70] Luca Trevisan. Extractors and pseudorandom generators. *Journal of the ACM*, 48(4):860–879, 2001.   17, 18

[71] Luca Trevisan. List-decoding using the XOR Lemma. In *Proceedings of the 44th IEEE Symposium on Foundations of Computer Science*, pages 126–135, 2003.   16

[72] Luca Trevisan. Some applications of coding theory in computational complexity. *Quaderni di Matematica*, 13:347–424, 2004. arXiv:cs.CC/0409044.   14

[73] Luca Trevisan. On uniform amplification of hardness in NP. In *Proceedings of the 37th ACM Symposium on Theory of Computing*, pages 31–38, 2005.   16

[74] Luca Trevisan and Salil Vadhan. Pseudorandomness and average-case complexity via uniform reductions. In *Proceedings of the 17th IEEE Conference on Computational Complexity*, pages 129–138, 2002.   16

[75] Christopher Umans. Pseudo-random generators for all hardnesses. *Journal of Computer and System Sciences*, 2(67):419–440, 2003.   18

[76] Salil Vadhan. Randomness extractors and their many guises. In *Proceedings of the 43rd IEEE Symposium on Foundations of Computer Science*, pages 9–10, 2002.   9

[77] Salil P. Vadhan. Constructing locally computable extractors and cryptosystems in the bounded-storage model. *Journal of Cryptology*, 17(1):43–77, 2004.   10

[78] Umesh Vazirani. *Randomness, Adversaries and Computation*. PhD thesis, University of California, Berkeley, 1986.   8

[79] Umesh Vazirani and Vijay Vazirani. Random polynomial time is equal to slightly random polynomial time. In *Proceedings of the 26th IEEE Symposium on Foundations of Computer Science*, pages 417–428, 1985.   8, 9

[80] Umesh V. Vazirani. Strong communication complexity or generating quasirandom sequences form two communicating semi-random sources. *Combinatorica*, 7(4):375–392, 1987.   8

[81] Emanuele Viola. The complexity of constructing pseudorandom generators from hard functions. *Computational Complexity*, 13(3-4):147–188, 2004.   16

[82] John von Neumann. Various techniques used in connection with random digits. *National Bureau of Standards, Applied Mathematics Series*, 12:36–38, 1951.   8

[83] Avi Wigderson and David Zuckerman. Expanders that beat the eigenvalue bound: Explicit construction and applications. *Combinatorica*, 19(1):125–138, 1999.   9

[84] Andrew C. Yao. Theory and applications of trapdoor functions. In *Proceedings of the 23th IEEE Symposium on Foundations of Computer Science*, pages 80–91, 1982. 3, 12, 13, 14

[85] Richard Zippel. Probabilistic algorithms for sparse polynomials. In *Proceedings of the International Symposiumon on Symbolic and Algebraic Computation*, pages 216 – 226. Springer-Verlag, 1979. 2

[86] David Zuckerman. General weak random sources. In *Proceedings of the 31st IEEE Symposium on Foundations of Computer Science*, pages 534–543, 1990. 4, 9

[87] David Zuckerman. Linear degree extractors and the inapproximability of max clique and chromatic number. Technical Report TR05-100, Electronic Colloquium on Computational Complexity, 2005. 8